\documentclass[runningheads]{llncs}
\usepackage{graphicx}

\usepackage{tikz}
\usepackage{comment}
\usepackage{amsmath,amssymb} 
\usepackage{color}
\usepackage{rotating}
\usepackage{multirow}
\usepackage{colortbl}
\usepackage[accsupp]{axessibility}  

\begin{document}
\pagestyle{headings}
\mainmatter
\def\ECCVSubNumber{100}  

\title{Contour Dice loss for structures with Fuzzy and Complex Boundaries in Fetal MRI} 

\titlerunning{Abbreviated paper title}
%
\author{Bella Specktor Fadida\inst{1} \and
Bossmat Yehuda\inst{2,3}\and
Daphna Link Sourani \inst{2} \and 
Liat Ben Sira \inst{3,4} \and
Dafna Ben Bashat \inst{2,3} \and Leo Joskowicz \inst{1}
}
\authorrunning{F. Author et al.}
%
\institute{School of Computer Science and Engineering, The Hebrew University of Jerusalem, Israel \and
Sagol Brain Institute, Tel Aviv Sourasky Medical Center, Israel \and
Sackler Faculty of Medicine \& Sagol School of Neuroscience, Tel Aviv University, Israel \and
Division of Pediatric Radiology, Tel Aviv Sourasky Medical Center, Tel Aviv-Yafo, Israel\\
\email{\{bella.specktor@mail.huji.ac.il,josko@cs.huji.ac.il\}}}
\maketitle

\begin{abstract}
Volumetric measurements of fetal structures in MRI are time consuming and error prone and therefore require automatic segmentation. Placenta segmentation and accurate fetal brain segmentation for gyrification assessment are particularly challenging because of the placenta fuzzy boundaries and the fetal brain cortex complex foldings. In this paper, we study the use of the Contour Dice loss for both problems and compare it to other boundary losses and to the combined Dice and Cross-Entropy loss. The loss is computed efficiently for each slice via erosion, dilation and XOR operators. We describe a new formulation of the loss akin to the Contour Dice metric. The combination of the Dice loss and the Contour Dice yielded the best performance for placenta segmentation. For fetal brain segmentation, the best performing loss was the combined Dice with Cross-Entropy loss followed by the Dice with Contour Dice loss, which performed better than other boundary losses. 

\end{abstract}

\section{Introduction}

Fetal MRI has the potential to complement ultrasound (US) imaging and improve fetal development assessment by providing accurate volumetric measurements of the fetal structures \cite{Reddy2008,Rutherford2008}. Since manual segmentation of the fetal structures is very time consuming and impractical, automatic segmentation is required. However, placental structure and fetal brain have fuzzy and complex contours, hence using standard loss functions may result in inaccuracies on their boundaries even if they demonstrate good performance for the segmentation of the general structure.

Recent works propose to address segmentation contour inaccuracies by using additional boundary and contour-based loss functions to train convolutional neural networks (CNN) \cite{Caliva2019,Jurdi2021,Karimi2019,Kervadec2019a,Specktor-Fadida2021,Yang2019}. Contour-based losses aim to minimize directly or indirectly the one-to-one correspondence between points on the predicted and label contour. Therefore, these losses are often complex and their computation cost is high. Recently, Jurdi et al \cite{Jurdi2021} proposed a computationally efficient loss that optimizes the mean squared error between the predicted and the ground-truth perimeter length. However, a question remains whether the global computation of the contour perimeter is sufficient to regularize very complex boundaries, e.g., the fetal brain contour. Specktor-Fadida et al \cite{Specktor-Fadida2021} proposed a loss function based on the contour Dice metric which performs local regularization of the contour and is efficient. It was reported to perform well on placenta segmentation, but it is not clear how it compares to other boundary losses. Moreover, it was not tested for other structures, e.g. the fetal brain.

Isensee et al \cite{Isensee2021} showed that the nnU-Net trained with the Dice loss function in combination with cross-entropy loss surpassed existing approaches on 23 public datasets with various segmentation structures and modalities. Despite the evident success of the compound Dice and Cross-Entropy loss, most of the works on contour losses compare results only to other boundary losses. Ma et al \cite{Ma2021} compared multiple compound loss functions, including the combination of Dice loss with boundary losses, i.e., the Hausdorff distance loss and the Boundary loss, and the combination of Dice loss with regional losses like the Cross Entropy and TopK losses. They reported that the Hausdorff loss yielded the best results for liver tumors segmentation. However, they did not test the losses for structures with fuzzy or complex boundaries. Also, since the paper was published, two additional boundary-based losses, the Perimeter loss and the Contour Dice loss, were introduced \cite{Jurdi2021,Specktor-Fadida2021}.

In this paper, we investigate the performance of the Contour Dice loss and other losses for the placenta and fetal brain segmentation in MRI with emphasis on their boundaries and describe a new formulation for the Contour Dice loss function. The main contributions are:  1) a quantitative comparison of the performance of the Contour Dice loss, of other state-of-the-art boundary losses and of the combined Dice and Cross-Entropy loss for the placenta and fetal brain segmentation; 2) a new formulation for the contour Dice loss which is more similar to the Contour Dice metric;  3) quantification of the effect of the contour extraction thresholding on both the Contour Dice and the Perimeter losses.

\section{Background and Related Work}

\subsection{Placenta Segmentation}
The placenta plays an important role in fetal health, as it regulates the transmission of oxygen and nutrients from the mother to the fetus. Placental volume is an important parameter to assess fetal development and identify cases at risk with placental insufficiency \cite{salavati2019}. Automatic placenta segmentation in fetal MRI poses numerous challenges. These include MRI related challenges, e.g., varying resolutions and contrasts, intensity inhomogeneity, image artifacts due to the large field of view, and partial volume effect, and fetal scanning related challenges, e.g., motion artifacts due to fetal and maternal movements, high variability in the placenta position, shape, appearance, orientation and fuzzy boundaries between the placenta and uterus.

Most of the existing methods for automatic placenta segmentation are based on deep learning. Alansary et al. \cite{Alansary16} describe the first automatic placenta segmentation method. It starts with a coarse segmentation with a 3D multi-scale CNN whose results are refined with a dense 3D Conditional Random Field (CRF). It achieves a Dice score of 0.72 with 4-fold cross validation on 66 fetal MRI scans. Torrents-Barrena et al. \cite{Torrents-Barrena2019} present a method based on super-resolution and motion correction followed by Gabor filters based feature extraction and Support Vector Machine (SVM) voxel classification. It achieves a Dice score of 0.82 with 4-fold cross-validation on 44 fetal MRI scans. Han et al. \cite{Han2019} present a method based on a 2D U-Net trained on a small dataset of 11 fetal MRI scans. It achieves a mean IU score of 0.817. Quah et al \cite{Quah2021} compares various methods for placenta segmentation on 68 fetal MRI 3D T2* images. The best method in this study achieves a Dice score of 0.808 using a U-Net with 3D patches. Pietsch et al. \cite{pietsch2021} describe a network that uses 2D patches and yields a Dice score of 0.76, comparable to expert variability performance. Specktor-Fadida et al \cite{Specktor-Fadida2021} demonstrate good performance for placenta segmentation using a new Contour Dice loss in combination with Soft Dice loss that yields a Dice score of 0.85.

\subsection{Fetal Brain Segmentation}
Fetal brain assessment, relating to its volume and structures, is important to assess fetal development and predict newborn outcome. MRI is often used as a complimentary tool mainly in cases with suspected brain malformations \cite{Hosny2010}. Automatic segmentation of the  fetal brain is important to assess changes of the brain volume with gestational age,  and accurate contour segmentation is necessary to assess the cortical folding of the brain, which was found to be an important biomarker for later functional development \cite{Dubois2008,Dubois2019}.

Multiple works were proposed for automatic fetal brain segmentation \cite{Dudovitch2020,Ebner2020,Salehi2018,torrents2019review}. Recently, a fetal brain tissue segmentation challenge was conducted for seven different fetal brain structures \cite{Payette2022}. However, most studies focus on fetal brain segmentation to assess the volume or its structure. In order to assess cortical folding sulcation, a more accurate segmentation of the outer contour is needed.

\subsection{Boundary Loss Functions}
A variety of papers propose to add a constraint to the loss function to improve the accuracy of the segmentation in the structure boundaries. Arif et al.~\cite{Arif2018} extend the standard cross-entropy term with an average point to curve Euclidean distance factor between predicted and ground-truth contours. This allows the network to take into consideration shape specifications of segmented structures. Caliva et al. \cite{Caliva2019} use distance maps as weighing factors for a cross-entropy loss term to improve the extraction of shape bio-markers and enable the network to focus on hard-to-segment boundary regions. As a result, their approach emphasizes the voxels that are in close proximity of the segmented anatomical objects over those that are far away. Yang et al. \cite{Yang2019} use Laplacian filters to develop a boundary enhanced loss term that invokes the network to generate strong responses around the boundary areas of organs while producing a zero response for voxels that are farther from the structures periphery. 

\begin{figure}
\centering
\includegraphics[height=3.4cm, width=12cm]{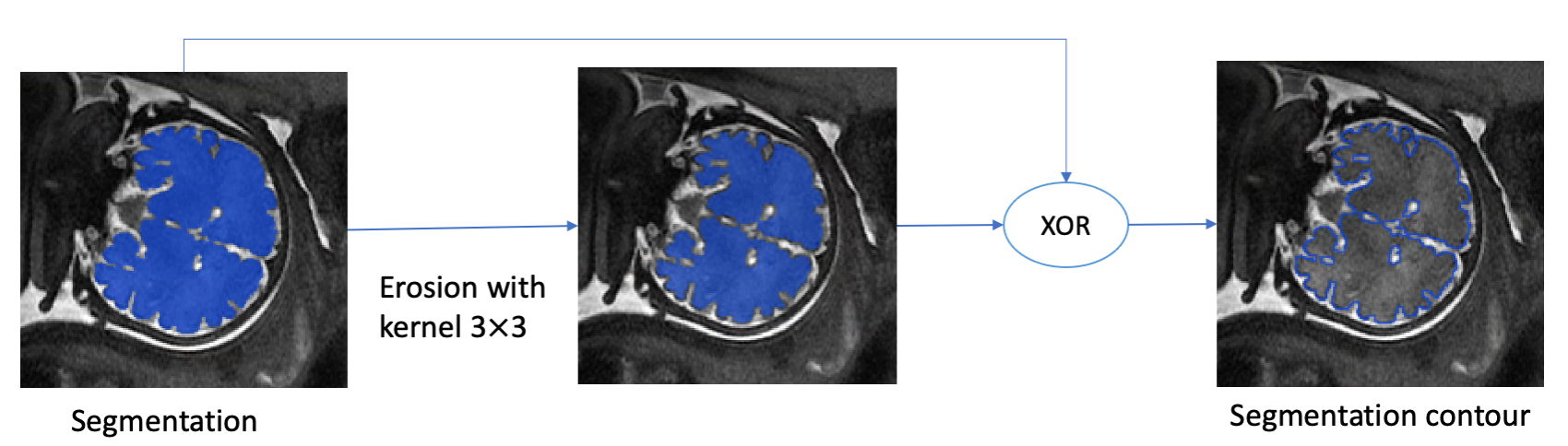}
\caption{Illustration of the contour extraction method using erosion and XOR operators. Erosion with a $3\times 3$ kernel produces a segmentation contour with a width of $2$ pixels.}
\label{fig:contour_extraction}
\end{figure}

Both the boundary loss \cite{Kervadec2019a} and the Hausdorff loss \cite{Karimi2019} aim to minimize the distance between the boundaries of the ground truth and the computed segmentation. The boundary loss approximates the distance between the contours by computing the integrals over the interface between the two boundaries mismatched regions. The Hausdorff loss, computed with the distance transform, approximates the minimization of the Hausdorff distance between the segmentation and ground truth contours. The Hausdorff loss was reported to perform relatively well compared to multiple other losses on four different segmentation tasks and yielded the best results for liver tumors segmentation \cite{Ma2021}. However, the loss is computationally costly due to the computation of the distance transform maps, especially for large 3D blocks.

Recently, two loss functions based on the segmentation contour were proposed. One is the Perimeter Loss \cite{Jurdi2021}, which optimizes the perimeter length of the segmented object relative to the ground-truth segmentation using the mean squared error function. Soft
approximation of the contour of the probability map is performed by specialized, non-trainable layers in the network. The second is the Contour Dice \cite{Specktor-Fadida2021}, based on the Contour Dice metric, which was shown to be highly correlated to the time required to correct segmentation errors \cite{Kiser2021,Nikolov2018}.

Formally, let $\partial T$ and $\partial S$ be the extracted surfaces of the ground truth delineation and the network results, respectively, and let $B_{\partial T}$ and $B_{\partial S}$ be their respective offset bands. The Contour Dice (CD) metric of the offset bands is:
\begin{align}
\label{eqn:contour_dice_metric}
CD = \frac{|\partial T\bigcap B_{\partial S}|+|\partial S\bigcap B_{\partial T}|}{|\partial T|+|\partial S|}
\end{align}

\begin{figure}[t]
\centering
\includegraphics[height=7cm, width=12cm]{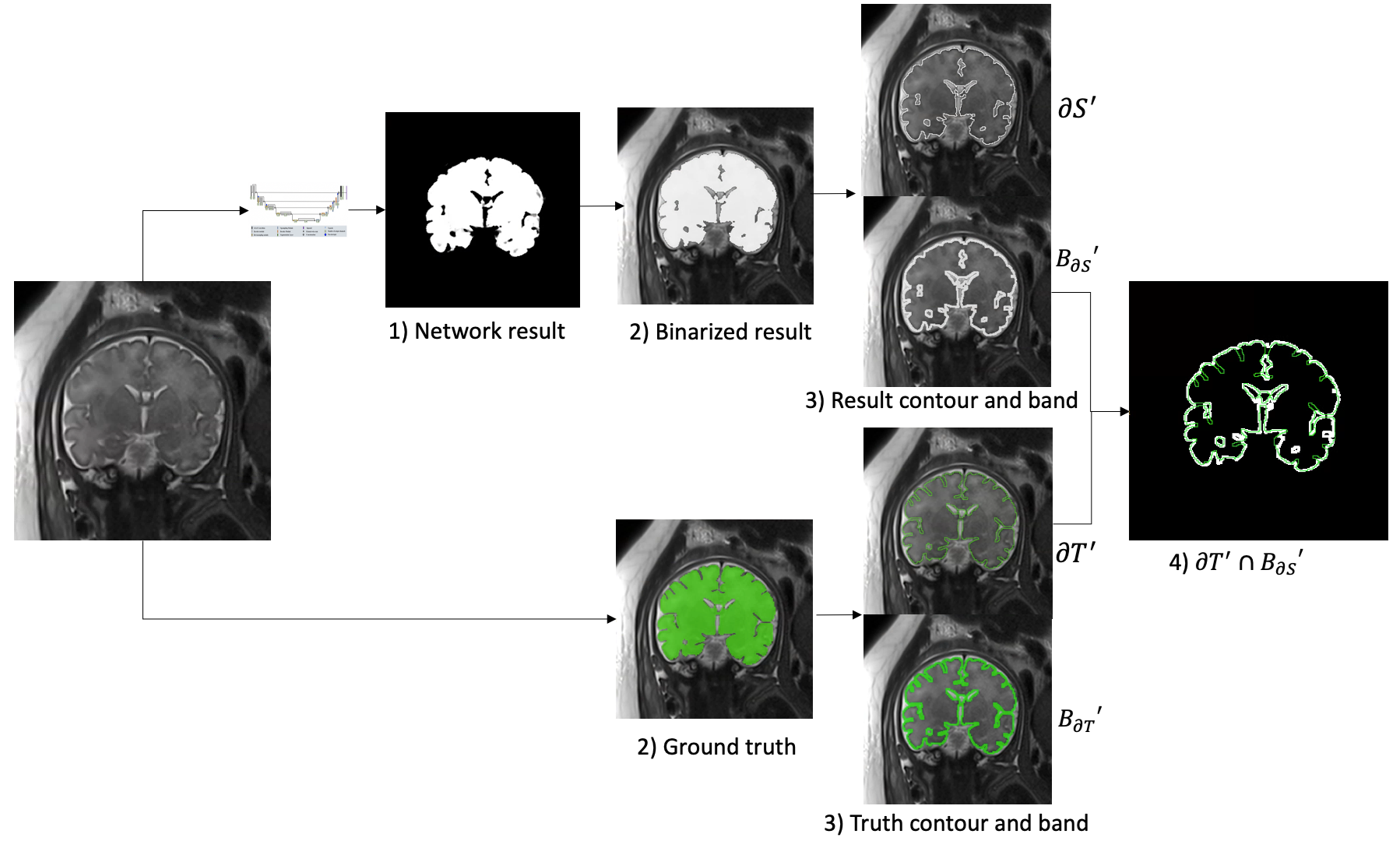}
\caption{Illustration of the intersection term computation in the Contour Dice loss. 1) Network result; 2) Binary masks of segmentation result and ground truth. To create a binary mask, thresholding is applied on the network result with threshold parameter $t$. The ground truth is already a binary mask; 3) Contours and bands are extracted; 4) Illustration of the computation of ${\partial T}'\bigcap {B_{\partial S}}'$. The light green are intersection areas, dark green are ground truth contour areas which do not intersect with the computed band and the white areas are  computed band areas which do not intersect with the ground truth contour. The computation of ${\partial S}'\bigcap {B_{\partial T}}'$ is performed similarly using the ${\partial S}'$ and ${B_{\partial T}}'$ masks.}
\label{fig:contour_dice}
\end{figure}

To make the Contour Dice loss function differentiable, Specktor-Fadida et al \cite{Specktor-Fadida2021} proposed the following formulation:
\begin{align}
L_{CD} = -\frac{2|B_{\partial T}\bigcap B_{\partial S}|}{|B_{\partial T}|+|B_{\partial S}|}
\label{eqn:CD1_loss}
\end{align}
This formulation makes the function trainable, as we integrate over bands which have a predefined width.

\section{Methods}
For the Contour Dice loss function calculation, we estimate the segmentation contours $\partial T$ and $\partial S$ from Eq.~\ref{eqn:contour_dice_metric} with the erosion and XOR operators (Fig. ~\ref{fig:contour_extraction}). As a result, the computed contours have a width. We can now integrate over the contour voxels and formulate a loss function that is very similar to the original Contour Dice metric in Eq.~\ref{eqn:contour_dice_metric}.

Formally, let ${\partial T}'$ be the extracted ground truth contour $\partial T$, ${\partial S}'$ be the extracted segmentation result contour $\partial S$ and ${B_{\partial T}}'$ and ${B_{\partial S}}'$ be their respective bands. We formulate the Contour Dice loss $L_{CD}$ function as:
\begin{align}
\label{eqn:CD2_loss}
L_{CD} = -\frac{|{\partial T}'\bigcap {B_{\partial S}}'|+|{\partial S}'\bigcap {B_{\partial T}}'|}{|{\partial T}'|+|{\partial S}'|}
\end{align}
The contour Dice loss calculation is performed in two steps: 1) Contour and band extraction for ground truth segmentation and binarized segmentation result; 2) Dice with Contour dice loss computation. Fig.~\ref{fig:contour_dice} illustrates the Contour Dice computation using the extracted contours and bands. 

\begin{figure}[t]
\centering
\includegraphics[height=6.1cm, width=12cm]{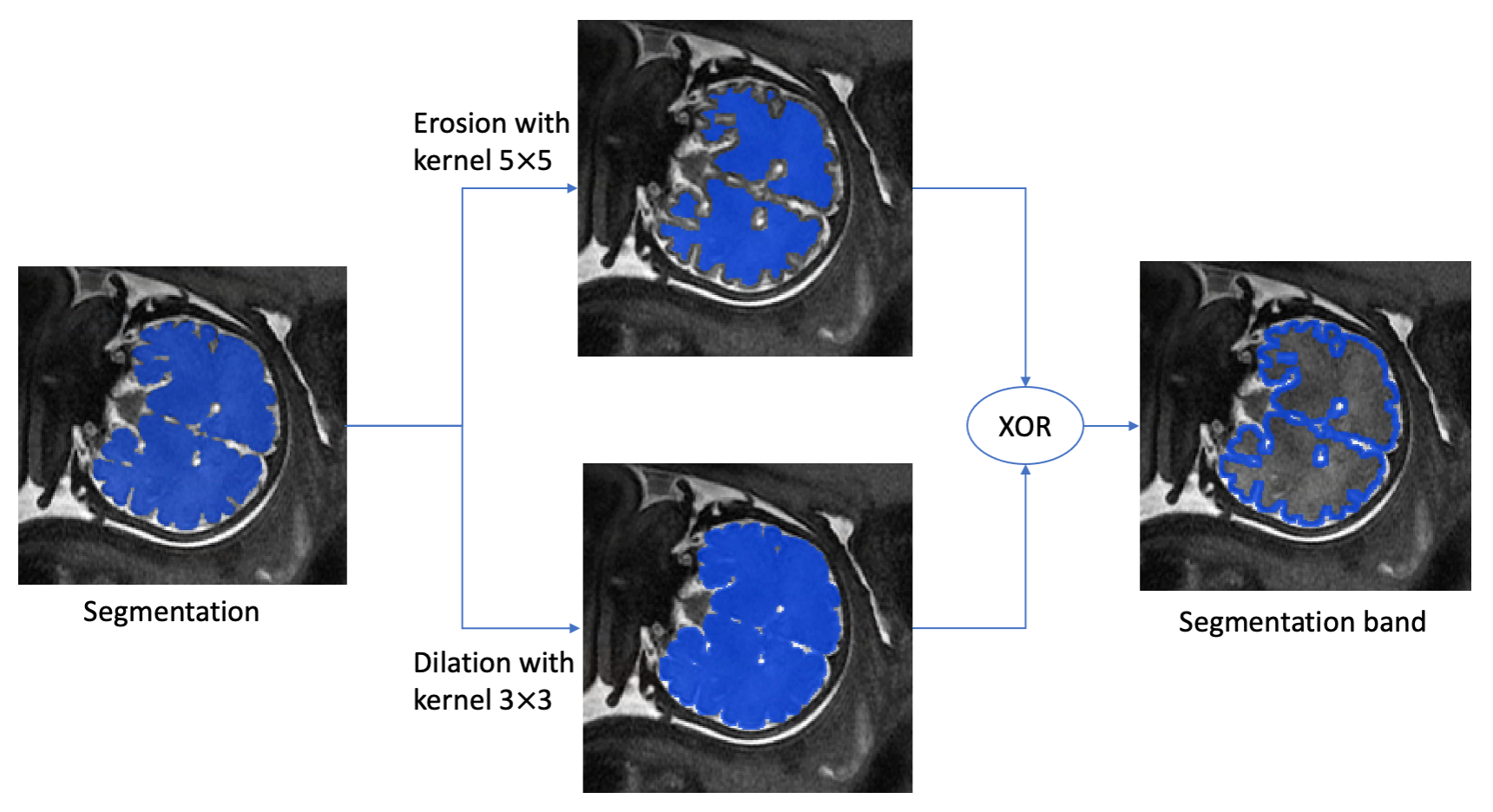}
\caption{Illustration of the segmentation offset band extraction method using erosion, dilation and XOR operators.}
\label{fig:band_extraction}
\end{figure}

\subsection{Contours and Bands Extraction} \label{Contour_extraction_method}
The segmentation contour and the band around it are computed as follows. First, binary thresholding is applied to the network output with a predefined threshold $t \in [0,1]$. Then, the contours of both the network result and the ground truth segmentation are extracted using erosion and XOR operations (Fig. \ref{fig:contour_extraction}). Finally, the bands are extracted using erosion, dilation and XOR operations (Fig.~ \ref{fig:band_extraction}).

\subsection{Contour Dice Loss Computation}
The smoothing term $\epsilon$ is added to the loss function formulation, similar to the Soft Dice:
\begin{align}
\label{eqn:CD2_loss_smoothing}
L_{CD} = -\frac{|{\partial T}'\bigcap {B_{\partial S}}'|+|{\partial S}'\bigcap {B_{\partial T}}'|+ \epsilon}{|{\partial T}'|+|{\partial S}'| + \epsilon }
\end{align}
The contour dice loss $L_{CD}$ in Eq.~\ref{eqn:CD2_loss_smoothing} is used in combination with the Soft Dice $L_{D}$ loss with a constant weighting scheme:
\begin{align}
\label{eqn:DCD_loss}
L = L_{D}+\gamma L_{CD}
\end{align}
We use the batch implementation for the Soft Dice and Contour Dice loss functions. There, the loss is simultaneously computed for all voxels in a mini-batch instead of a separate computation for each image \cite{Kodym2018}.

\section{Experimental Results}
To evaluate our method, we retrospectively collected fetal MRI scans for placenta segmentation and for fetal brain segmentation. 

\subsection{Datasets and Annotations}
We collected fetal MRI scans acquired with the FIESTA, FRFSE and HASTE sequences from the Sourasky Medical Center (Tel Aviv, Israel). 
Gestational ages (GA) were 22-39 weeks, with most cases 28-39 weeks. For placental segmentation, FIESTA dataset was used, consists of 40 labeled cases acquired on a 1.5T GE Signa Horizon Echo speed LX MRI scanner using a torso coil. Each scan has 50–100 slices, 256×256 pixels/slice, and resolution of 1.56×1.56×3.0 $mm^3$. For fetal brain segmentation, the dataset consists of 48 FRFSE and 21 HASTE cases. The FRFSE cases were acquired on 1.5T MR450 GE scanner. The HASTE cases were acquired on 3T Skyra and Prisma Siemens scanners. Each scan had 11-46 slices, 49-348×49-348 pixels/slice and resolution of 0.40-1.25×0.40-1.25×2.2-6 $mm^3$.

Ground truth segmentations were created as follows. For placenta segmentation cases, 31 cases were annotated from scratch and 9 additional cases were manually corrected from the network results. For brain segmentation cases, all cases were annotated by correcting network results. Both the annotations and the corrections were performed by a clinical expert.

\subsection{Experimental Studies}
We conducted three studies. The first two studies compare different losses performance: Dice loss, Dice with Cross-Enropy loss, Dice with Boundary loss, Dice with Perimeter loss, Dice with Hausdorff loss and Dice with Contour Dice loss. Study 1 compares the losses for the task of placenta segmentation and Study 2 compares the losses for the task of fetal brain segmentation. Study 3 quantifies the influence of the segmentation result threshold parameter in the contour extraction phase on two contour-based losses, the Perimeter loss and the Contour Dice loss.

In all experiments, the Contour Dice loss weight from Eq.~\ref{eqn:DCD_loss} was set to a constant $\gamma=0.5$; the weight of Cross Entropy loss was set to $1$, i.e. the same weight for Dice and Cross-Entropy terms. The weight for the other boundary losses, the Boundary loss, the Perimeter loss and the Hausdorff loss, were set with the dynamic scheme in \cite{Jurdi2021,Kervadec2019a}. The schedule was as follows: the initial weight was set to 0.01 and increased by 0.01 every epoch. The network was trained with reducing learning rate on plateau with early stopping of 50 epochs in case there is no reduction of the loss on the validation set. 

We used the contour extraction method in Sec.~\ref{Contour_extraction_method} for the Perimeter and Contour Dice losses. Since ground truth segmentations for both placenta and fetal brain structures are of a high quality, the band for the Contour Dice loss in all experiments was set to be the estimated contour ${B_{\partial T}}'={\partial T}'$ and ${B_{\partial S}}'={\partial S}'$ (Eq.~\ref{eqn:CD2_loss_smoothing}).

A network architecture based on Dudovitch et al \cite{Dudovitch2020} was used in all experiments. For placenta segmentation we used patch size of 128×128×48 and for fetal brain segmentation we used a large patch size of 256×256×32 because of a higher in-plane resolution. The segmentation results were refined by standard post-processing techniques. Hole filling was used only for the placenta segmentation and not for the fetal brain because the brain structure does have holes. 

Prior to the use of the the placenta segmentation network, 
the same detection network was used in all experiments to extract the Region of Interest (ROI) around the placenta. The detection network architecture was similar to \cite{Dudovitch2020} and was applied on down-scaled scans by $ \times 0.5$ in the in-plane axes. Since the fetal brain dataset was constructed after brain ROI extraction, there was no need to apply a fetal brain detection network.

Segmentation quality was evaluated in all studies with the Dice, Hausdorff and 2D ASSD (slice Average Symmetric Surface Difference) metrics. 3D ASSD was not evaluated as it is very dependent on the surface extraction method.

\begin{table}[t]
\caption{Placenta segmentation results comparison.}
\label{table:placenta_results}
\centering
\begin{tabular}{|l||c|c|c|} 
\hline
      LOSS/METRIC               & Dice                 & Hausdorff            & ASSD              \\ 
\hline
Dice                 & 0.773±0.117          & 57.73±44.24          & 8.35±7.43          \\ 
\hline
Dice + Cross Entropy & 0.807±0.098          & 50.48±40.15          & 5.83±3.34           \\ 
\hline
Dice + Boundary      & 0.805±0.079          & 52.78±40.68          & 6.76±3.54           \\ 
\hline
Dice + Perimeter     & 0.817±0.061          & 50.84±39.32          & 5.92±1.88           \\ 
\hline
Dice + Hausdorff     & 0.829±0.069          & 49.40±43.28          & 5.63±4.56           \\ 
\hline
Dice + Contour Dice  & \textbf{0.847±0.058} & \textbf{44.60±42.31} & \textbf{4.46±2.45}  \\
\hline
\end{tabular}
\end{table}

\begin{figure}[t]
\centering
\includegraphics[height=5cm, width=12cm]{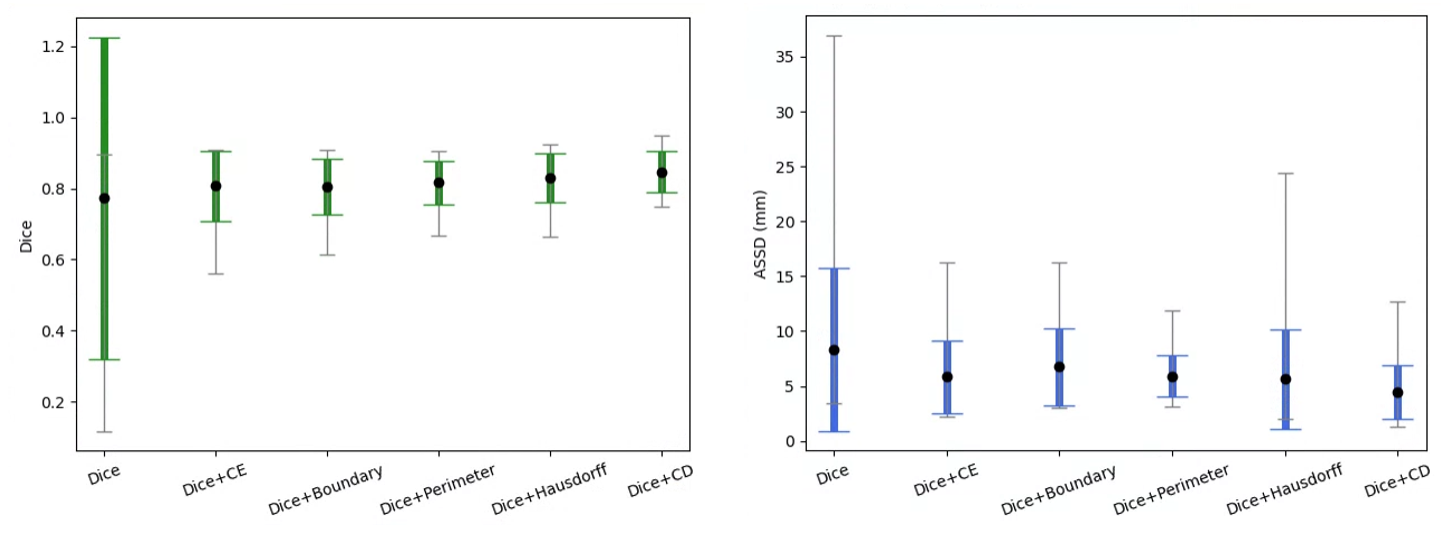}
\caption{Placenta segmentation results bar plots comparison for the Dice score (left, green) and 2D ASSD (right, blue) metrics. The color bars show the Standard Deviation. The gray bars show minimum and maximum values. X-axis from left to right: Dice loss, Dice with Cross-Entropy loss, Dice with Boundary loss, Dice with Perimeter loss, Dice with Hausdorff loss and Dice with Contour Dice loss.}
\label{fig:placenta_plot}
\end{figure}

\begin{figure}[t]
\centering
\includegraphics[height=7cm, width=12cm]{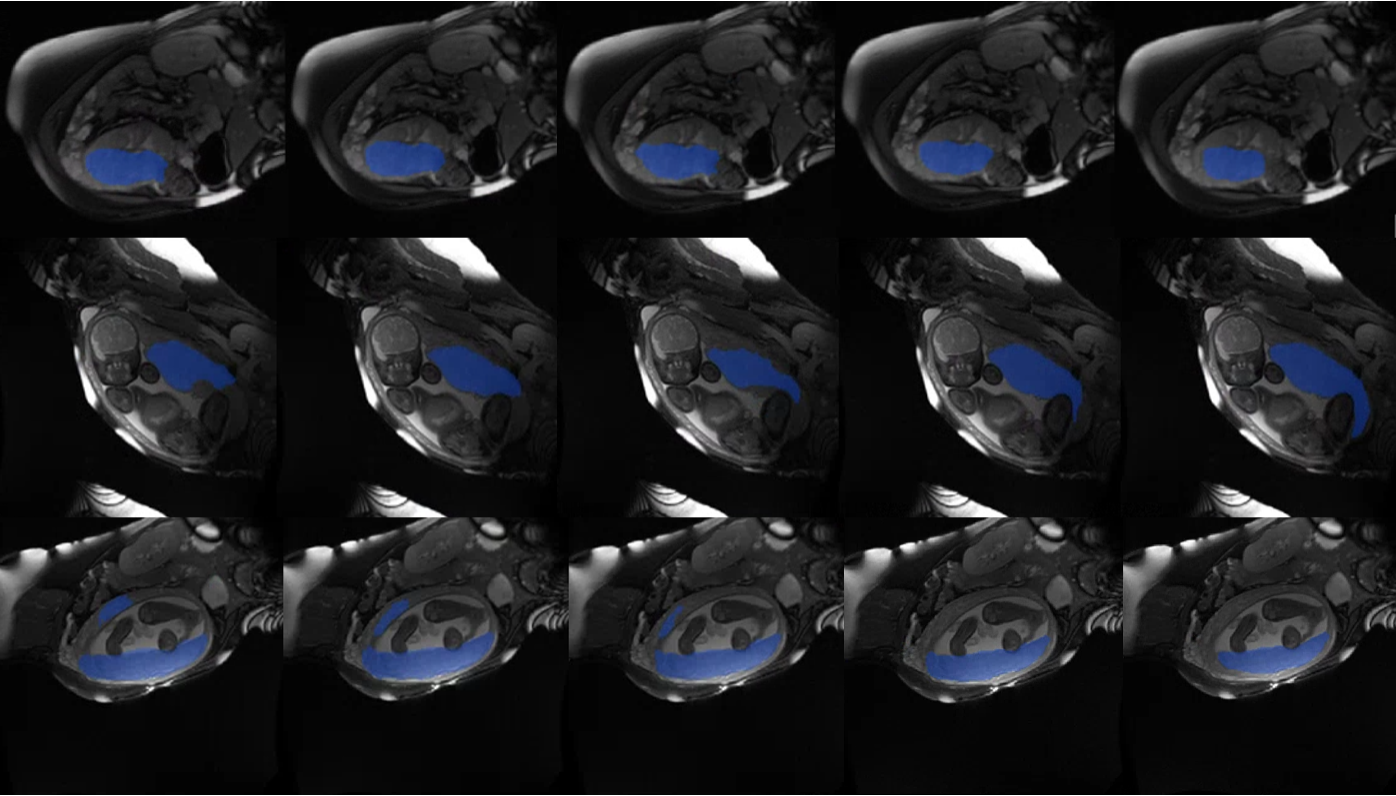}\\
(a) \hspace*{20mm} (b)\hspace*{20mm} (c)\hspace*{20mm} (d)\hspace*{20mm} (e)
\caption{Illustrative placenta segmentation results. (a) Dice and Boundary loss; (b) Dice and Perimeter loss; (c) Dice and Hausdorff loss; (d) Dice with contour Dice loss; (e) ground truth segmentation}
\label{fig:placenta_results}
\end{figure}

\subsubsection{Study 1: placenta segmentation}
The Contour Dice loss method was evaluated on training, validation and test sets of 16, 3 and 21 annotated placenta FIESTA scans respectively and compared to other state-of-the art losses. The segmentation threshold was set to $t=1$ for both the Perimeter and the Contour Dice losses, as it yielded better results than $t=0.5$, with a significant performance difference for the Contour Dice loss (see Study $3$ for more details). 

Table~\ref{table:placenta_results} lists the results of various loss functions for placenta segmentation. All compound losses improved upon the Dice loss alone, with the combination of Dice with Contour Dice loss yielding the best Dice score of 0.847. The second best performance was of the combination of Dice with the Hausdorff loss, with a Dice score of 0.829, better than the highly used Dice with Cross-Entropy loss combination. The combination of Dice loss with the Perimeter loss also yielded an improvement upon the Dice with Cross-Entropy loss.

Fig.~\ref{fig:placenta_results} shows illustrative examples of placenta segmentation results with the boundary-based losses. The Contour Dice loss yields a smooth and relatively accurate segmentation. The perimeter loss also yields a smooth segmentation but sometimes fails to capture the full placenta shape complexity. The Hausdorff loss captures well the placenta shape, but it sometimes misses on the boundary.

\subsubsection{Study 2: fetal brain}
The Contour dice loss method was compared to the other state-of-the-art losses on the brain segmentation task using
35/4/30 training/validation/test split. We set the segmentation threshold of $t=1$ for the Contour Dice loss, and set $t=0.5$ for the Perimeter loss as it performed significantly better (see Study 3 for more details). All losses were trained with an early stopping of 50 epochs except the Hausdorff loss, which was trained for 42 epochs because of slow training on the large block size of 256×256×32 and GPU cluster time constraints.

Table~\ref{table:brain_results} lists the results of various loss functions for fetal brain segmentation. Surprisingly, the best performing loss was not a boundary-based loss, but the combination of the Dice loss with the Cross-Entropy loss, with a Dice score of 0.954 and 2D ASSD of 0.68 mm. The combination of the Dice loss with the Contour Dice had slightly worse performance with a Dice score of 0.946 and 2D ASSD of 0.81 mm. This loss performed best compared to all other boundary losses, with the Hausdorff loss having the third best performance with a Dice score of 0.932 and 2D ASSD of 0.93 mm.

\begin{table}[t]
\centering
\caption{Brain segmentation results comparison.}
\label{table:brain_results}
\begin{tabular}{|l|c|c|c|} 
\hline
          LOSS/METRIC            & Dice                 & ~Hausdorff~        & ASSD                \\ 
\hline
Dice                  & 0.924±0.035          & 11.05±4.57         & 1.12±0.54           \\ 
\hline
Dice + cross-entropy~ & \textbf{0.954±0.019} & \textbf{7.65±1.86} & \textbf{0.68±0.26}  \\ 
\hline
Dice + boundary       & 0.924±0.039          & 10.45±4.28         & 1.16±0.57           \\ 
\hline
Dice + perimeter      & 0.932±0.032          & 11.07±3.27         & 1.00±0.43           \\ 
\hline
Dice + Hausdorff      & 0.934±0.032          & 10.17±2.83         & 0.93±0.36           \\ 
\hline
Dice + contour Dice   & 0.946±0.026          & 8.24±2.16          & 0.81±0.35           \\
\hline
\end{tabular}
\end{table}

\begin{figure}[t]
\centering
\includegraphics[height=5cm, width=12cm]{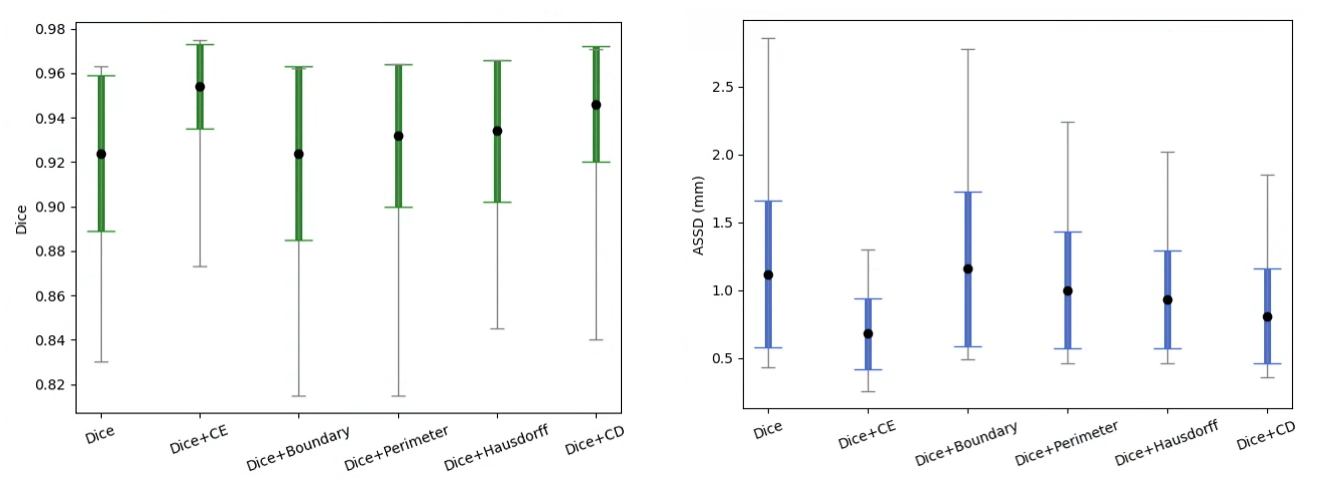}
\caption{Fetal brain segmentation results bar plots comparison for the Dice score (left in green) and 2D ASSD (right in blue) metrics. The color bars show the Standard Deviation. The gray bars show minimum and maximum values. X-axis from left to right: Dice loss, Dice with Cross-Entropy loss, Dice with Boundary loss, Dice with Perimeter loss, Dice with Haudorff loss and Dice with Contour Dice loss.}
\label{fig:cortex_plot}
\end{figure}

\begin{figure}[t]
\centering
\includegraphics[height=5cm, width=12cm]{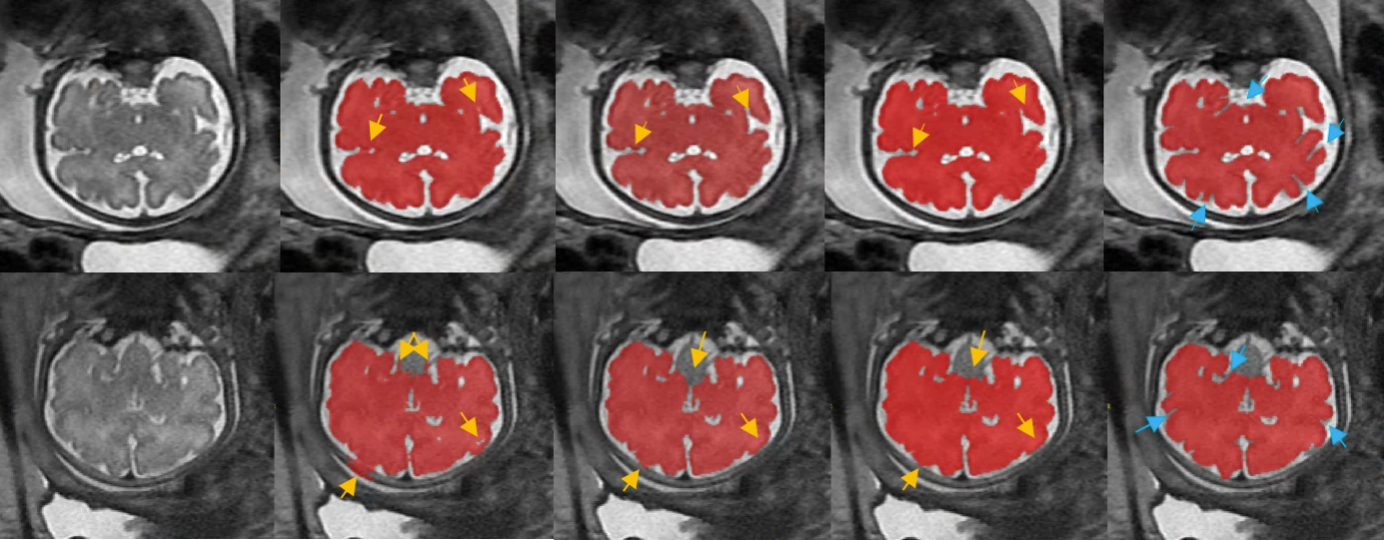}\\
(a) \hspace*{20mm} (b)\hspace*{20mm} (c)\hspace*{20mm} (d)\hspace*{20mm} (e)
\caption{Illustrative brain segmentation results: (a) original scan; (b) Dice loss; (c) Dice and Contour Dice loss; (d) Dice and Cross-Entropy loss; (e) ground truth segmentation. Yellow arrows point to segmentation errors of the Dice loss and their correction using either the Dice with Contour Dice or Dice with Cross-Entropy loss. The blue arrows point to cortex sulcation regions that were missed by all losses.}
\label{fig:brain_results}
\end{figure}

\begin{table}[t]
\centering
\caption{Ablation results for values of the segmentation threshold $t$ for contour extraction for the Dice with Contour Dice loss (DCD) and Dice with Perimeter loss (DP).}
\label{table:threshold_ablation}
\begin{tabular}{|l||c|c|c|c||c|c|c|} 
\hline
\multicolumn{1}{|c|}{\begin{sideways}\end{sideways}} & \multirow{2}{*}{$t$} & \multicolumn{3}{c|}{\textbf{Placenta}}      & \multicolumn{3}{|c|}{\textbf{Fetal Brain}}  \\ 
\cline{3-8}
                                                     &                    & ~ ~Dice~ ~     & ~ ~Hausdorff~~ & ~ ASSD~~      & ~ Dice~~       & ~Hausdorff~    & ASSD               \\ 
\hline
\multirow{2}{*}{\textbf{DCD}}                                 & ~0.5~              & 0.798          & 52.04          & 6.85          & 0.939          & 9.22           & 0.90               \\ 
\cline{2-8}
                                                     & 1                  & \textbf{0.847} & \textbf{44.60} & \textbf{4.46} & \textbf{0.946} & \textbf{8.24}  & \textbf{0.82}      \\ 
\hline \hline
\multirow{2}{*}{\textbf{DP}}                                  & 0.5                & 0.813          & \textbf{48.45} & 6.09          & \textbf{0.932} & \textbf{11.07} & \textbf{1.00}      \\ 
\cline{2-8}
                                                     & 1~                 & \textbf{0.817} & 50.85          & \textbf{5.92} & 0.895          & 17.19          & 1.59               \\
\hline
\end{tabular}
\end{table}

Fig.~\ref{fig:brain_results} shows illustrative results of fetal brain segmentations. While the combined Dice loss and Cross Entropy or Contour Dice losses improved upon the segmentation results of the Dice loss alone, neither captured the full complexity of the brain cortex sulcation.

\subsubsection{Study 3: effect of the contour extraction threshold value.}\label{thresholding_study}
We examined the effect of the threshold parameter $t$ for contour extraction on the Perimeter and the Contour Dice losses.  All experimental conditions except for the value of $t$ remained the same. Since the threshold $t$ roughly captures the uncertainty of the network in the output prediction: $t=0.5$ means that the network has a certainty of $50\%$, while values closer to $t=1$ mean that the network has a very high certainty about the prediction.

Table~\ref{table:threshold_ablation} lists the results. For the Perimeter loss, the best performing threshold for fetal brain segmentation task was $t=0.5$, significantly improving performance compared to a threshold of $t=1$ from a Dice score of 0.895 to 0.932. For placenta segmentation there was no significant difference between the two. For the Contour Dice loss, the opposite was true. A threshold of $t=1$ resulted in better performance for both placenta and fetal brain segmentation tasks, improving placenta segmentation from a Dice score of 0.798 to 0.847 and fetal brain segmentation from 0.939 to 0.946.

A looming question is why the difference between the two losses is happening. We hypothesize that for the Perimeter loss it is important to capture the full length of the contour even if the output prediction is not certain because it is a global length property. On the other hand, the Contour Dice loss is applied locally, and it may be beneficial to push the network toward certain contour regions.

\section{Conclusions}
This paper demonstrates the effectiveness of the Contour Dice loss compared to other boundary losses and the combined Dice with Cross-Entropy loss for placenta segmentation, a structure with fuzzy boundaries and high manual delineation variability, and for fetal brain segmentation, a structure that has complex boundaries with gyrus and sulcus. 

We presented a new formulation for the Contour Dice loss which is more similar to the Contour Dice metric that was found to highly correlate with segmentation correction time. This formulation may be useful for cases with inaccurate boundaries of the ground truth segmentation as it poses relaxation on the contours intersection term.

For contour extraction, we performed thresholding on the segmentation followed by and erosion and XOR operators. We found that the segmentation performance of both the Contour Dice and the Perimeter losses was sensitive to the thresholding parameter $t$. Future work can explore other contour extraction techniques, i.e., min and max pooling operators as proposed in \cite{Jurdi2021}.

While the combined Dice with Contour Dice loss resulted in best performance for the task of placenta segmentation, it demonstrated only the second best performance for the task of brain segmentation, with best performance achieved by the compound Dice with Cross-Entropy loss. Maybe a better parameters tuning, a different optimization scheme or a combination with another regional loss function could improve the performance of boundary losses. In any case, the results show the importance of a comparison to the widely used Dice with Cross-Entropy loss even for tasks with fuzzy or complex contours where intuitively boundary losses should improve performance. 

\section*{Acknowledgement}
This research was supported in part by Kamin Grants 72061 and 72126 from the Israel Innovation Authority.

%
%
\bibliographystyle{splncs04}
\bibliography{egbib}
\end{document}